\documentclass[sigconf]{acmart}

\usepackage{subcaption}
\usepackage{multirow,graphicx}
\usepackage[dvipsnames]{xcolor} 
\usepackage[detect-all]{siunitx}
\usepackage{bm}
\usepackage{balance}

\usepackage{arydshln}
\usepackage{hyphenat}
\usepackage{array,csquotes}

\AtBeginDocument{%
  }

\copyrightyear{2026}
\acmYear{2026}
\setcopyright{cc}
\setcctype{by}
\acmConference[CIKM '26]{Proceedings of the 35th ACM International Conference on Information and Knowledge Management}{November 07--11, 2026}{Rome, Italy}
\acmBooktitle{Proceedings of the 35th ACM International Conference on Information and Knowledge Management (CIKM '26), November 07--11, 2026, Rome, Italy}
\acmDOI{10.1145/3799682.3840602}
\acmISBN{979-8-4007-2539-5/2026/11}

\newcommand{\up}{\bm{\scriptscriptstyle\uparrow}}
\newcommand{\down}{\bm{\scriptscriptstyle\downarrow}}

\begin{document}

\title[Investigating the Effects of Context on User Modeling with Feature-Conditioned Markov Models]{Beyond States: Investigating the Effects of Context on User Modeling with Feature-Conditioned Markov Models}

\author{Jana Isabelle Friese}
\orcid{0009-0005-2483-0476}
\affiliation{%
  \institution{University of Duisburg-Essen}
  \city{Duisburg}
  \country{Germany}
}
\email{jana.friese@uni-due.de}

\author{Andreas Konstantin Kruff}
\orcid{0009-0002-8350-154X}
\affiliation{%
  \institution{TH Köln - University of Applied Sciences}
  \city{Cologne}
  \country{Germany}
}
\email{andreas.kruff@th-koeln.de}

\author{Timo Breuer}
\orcid{0000-0002-1765-2449}
\affiliation{%
  \institution{TH Köln - University of Applied Sciences}
  \city{Cologne}
  \country{Germany}
}
\email{timo.breuer@th-koeln.de}

\author{Philipp Schaer}
\orcid{0000-0002-8817-4632}
\affiliation{%
  \institution{TH Köln - University of Applied Sciences}
  \city{Cologne}
  \country{Germany}
}
\email{philipp.schaer@th-koeln.de}

\author{Norbert Fuhr}
\orcid{0000-0002-0441-6949}
\affiliation{%
  \institution{University of Duisburg-Essen}
  \city{Duisburg}
  \country{Germany}
}
\email{norbert.fuhr@uni-due.de}

\renewcommand{\shortauthors}{Jana Isabelle Friese, Andreas Konstantin Kruff, Timo Breuer, Philipp Schaer, and Norbert Fuhr}

\begin{abstract}

User behavior simulation is widely used to evaluate interactive information retrieval systems, but classical state-based approaches (e.g., Markov models) have limited ability to incorporate contextual information relevant for decision-making. We address this limitation by introducing a feature-conditioned Markov-style user model, in which transition probabilities are modeled as functions of positional, content-based, and interaction-derived features, enabling context-aware decision making while preserving the structural simplicity and computational efficiency of state-based models.

Applying a multi-level framework that assesses predictive fit and behavioral fidelity, we analyze how different sources of contextual information contribute to realistic user simulation across multiple datasets, search settings, and feature configurations. Our results show that incorporating contextual features improves the models' ability to reproduce key aspects of real user interactions, but that their effectiveness hinges on search scenario and modeling objective. Instead of a \enquote{one-size-fits-all} solution, effective simulation requires task- and setting-specific feature selection. Our framework provides a practical and interpretable basis for making these choices.

\end{abstract}





\maketitle

\section{Introduction}

User behavior is central to the evaluation of interactive information retrieval (IIR) systems. While user studies provide valuable, user-centric insights~\cite{Kelly2009,Liu2019}, much of IR evaluation relies on system-oriented methods that abstract away from actual user interactions, and most common effectiveness measures implicitly assume particular models of user behavior, for example, with respect to attention, examination depth, or stopping behavior~\cite{Carterette2011,Jarvelin2002,Zhang2009}.

User simulation provides a principled way to make such behavioral assumptions explicit and to study their effects in a controlled, repeatable manner, thereby contributing to greater transparency and reproducibility in IIR~\cite{Balog2024}. By generating interaction sequences rather than relying solely on static relevance judgments, simulation-based evaluation can increase the realism of system-oriented experiments.
However, the usefulness of simulation critically depends on how well it reflects real user behavior: unrealistic simulators can result in incorrect conclusions about system performance~\cite{Azzopardi2016}.

A wide range of user models has been proposed for this purpose. Classical state-based approaches, such as Markov models, offer transparent and controllable representations of user behavior. They are computationally efficient, easy to analyze, and well-suited for systematic experimentation~\cite{Benoit2005,Yang2018}. However, these models typically assume context-free transitions, despite strong evidence that context substantially influences search behavior~\cite{Engelmann2024,Liu2017,White2009}. This limits their ability to capture important aspects of real user behavior.

Recent modeling approaches, such as large language models, can generate user behavior considering rich context and complex decision mechanisms~\cite{Park2023,Kruff2025}. Despite offering great flexibility, such approaches suffer from high computational cost and limitations in interpretability, systematic analysis, reproducibility, and control~\cite{Bommasani2022}. Moreover, in light of documented discrepancies between human and LLM relevance judgments, their validity as proxies for real user behavior remains uncertain~\cite{Alaofi2024}.
This highlights a methodological gap: there is currently no clear middle ground between simple, interpretable models and highly expressive, yet opaque ones.

In this work, we propose a principled extension of state-based user models that aims to bridge this gap. Instead of using constant transition probabilities, we model transitions as functions of contextual features. This allows decisions to depend not only on the current interaction state but also on various contextual factors.

This design makes it possible to investigate which contextual factors matter most for reproducing real user interactions across search settings, motivating the following research questions:

\begin{itemize}
    \item[RQ1] Does conditioning state-based user models on contextual features improve predictive fit and behavioral realism?
    \item[RQ2] In what ways do different types of contextual features, including situational and content-based information, contribute to reproducing different behavioral aspects? 
\end{itemize}

To answer these questions, we compare a baseline Markov model with a set of feature-conditioned variants and evaluate them using a multi-level evaluation framework that assesses both predictive performance and behavioral fidelity at varying levels of granularity across datasets capturing different search settings.

This work makes three main contributions: First, we introduce a framework for feature-conditioned state-based user modeling that combines interpretability with contextual expressiveness. Second, we present a structured evaluation methodology for assessing simulation quality across multiple levels of behavioral specificity. Third, we provide an empirical analysis across multiple datasets of how different types of contextual information affect simulated behavior, highlighting both where context improves realism and how the importance of specific features varies across settings.

\section{Related Work}

A widely adopted model-based approach for user modeling in information retrieval (IR) is the use of Markov models. Zerhoudi et al.~\cite{10.1145/3503162.3503165} argue that Markov models are popular for modeling user search behavior because of their simplicity, compact representation, and well-established theoretical foundations.

In particular, the authors highlight that first-order Markov models are well-suited for simulation-based modeling, as they capture essential sequential dependencies without introducing unnecessary complexity. Thus, here, Markov models are used for user modeling, with the model learned from logged user interaction data. Simulation, in contrast, refers to generating synthetic behavior by sampling from the learned model.~\cite{10.5555/537562}.

The simulation framework SimIIR 2.0 integrates Markov model\hyp{}based behavior modeling~\cite{10.1145/3511808.3557711}. Beyond first-order and k-th-order Markov models, several other Markov-based approaches have been employed in the IR context. These include Hidden Markov Models (HMMs)~\cite{DBLP:journals/tcdl/Dungs16} and Partially Observable Markov Decision Processes (POMDPs), which allow modeling latent user states and incomplete observability. Furthermore, decision-oriented frameworks such as Markov Decision Processes (MDPs) and Multi-Armed Bandits (MABs) have been successfully applied to capture interactive and adaptive retrieval scenarios~\cite{Yang2018}.
Building on these various Markov-based approaches, first-order Markov models with absorbing states were adopted for the experiments. This choice allows to capture sequential user navigation while explicitly modeling terminal behaviors such as session completion. The simplicity of first-order models ensures tractable estimation, while absorbing states provide a more realistic representation of user sessions~\cite{10.1007/978-3-319-69644-7_17}.
Despite the advantages, these models still rely on global transition probabilities, disregarding external factors. To address this limitation, context-aware Markov models have been proposed, allowing state transitions to depend on relevant contextual information and thereby improving predictive accuracy and descriptive power. Although contextual Markov models were already widely used in cybersecurity, Bar et al.~\cite{DBLP:journals/kbs/BarSR23} appear to be the first to apply this approach to user modeling in a web browsing scenario, representing URLs as states and transitions between them as actions. They propose different methods of incorporating context; however, only the Na\"ive-Bayes-style approach directly extends the Markov models with contextual features, representing them as discrete conditioning variables for look-up-based probability estimations. Building on this, our approach represents context as continuous feature vectors that parametrically shape transition probabilities.

In addition to modeling sequential navigation, understanding which signals influence user click behavior is crucial for improving and validating our approach. Previous studies have shown that users are strongly affected by position bias, tending to click more frequently on higher-ranked documents, while results lower in the ranking or on subsequent pages receive considerably less attention~\cite{10.1145/1229179.1229181,10.1145/1240624.1240691, DBLP:journals/jasis/LorigoHBXJGGPP08}. Conversely, some domain-specific relevance signals appear to have minimal impact on click decisions; for example, Kacem et al.~\cite{DBLP:conf/ecir/KacemM18} found that bibliometric indicators did not significantly influence user clicks. Beyond individual clicks, several studies have also highlighted that the overall quality of the result list can affect users’ stopping behavior, shaping how many results they examine and when they terminate their search session~\cite{10.1145/3498366.3505770, Azzopardi2020, 10.1145/3209978.3210075}.
Taken together, these findings suggest that both contextual factors and list-level signals shape user interactions, highlighting the need for models that can integrate multiple types of information when predicting click behavior and session outcomes.

\section{Proposed Model}

In this work, we aim to enhance Markov models by incorporating contextual information, and we evaluate how this affects the realism of simulated behavior. To this end, we consider two types of state-based user models, which we describe in the following sections.

\subsection{Base Model}

The base model serves as a baseline for comparison with our enhanced models. It is a standard Markov model with constant transition probabilities $P(a_{t+1} | s_t)$, where the state space is $S = \{s_1,...s_n\}$ and the set of possible actions is $A = \{a_1,...,a_m\}$. The probability of a particular action depends solely on the current state and is independent of any additional context.
While simple and interpretable, this approach cannot account for contextual factors such as document position, relevance, or the user’s prior interactions.

\subsection{Feature-conditioned Model}
To incorporate context, we propose an extension in which each state transition is modeled using a probabilistic classification model. Instead of fixed probabilities that depend only on the current state, the likelihood of each possible next action also depends on a feature vector $\boldsymbol{x}_t$ describing the current context at time $t$:
\[P(a_{t+1}|s_t,\boldsymbol{x}_t)\] 

By incorporating context, the model can adjust transition probabilities based on information available at each point in the session, capturing nuanced aspects of behavior beyond the current state.

Strictly speaking, these models are no longer Markovian, because some features implicitly encode session history, which violates the Markov property that the decision depends only on the present state $s_t$ alone. However, the feature-conditioned model can be interpreted as a first-order Markov model over an implicitly augmented state space. We define the augmented state as
\[\tilde{s}_t = (s_t,\boldsymbol{x}_t)\]
The transition probability then depends only on the current augmented state: 
\[ P(a_{t+1}|\tilde{s}_t)\] 

Conceptually, this corresponds to a Markov model with a very large, sparse transition matrix, where each combination of state and feature values is a separate state. While the explicit state space would be prohibitively large, the probabilistic classification parametrization provides a compact representation, allowing the model to remain interpretable and tractable. 

In summary, the proposed feature-conditioned model a) extends the base Markov model by conditioning transitions on contextual features, b) preserves the Markov property in an augmented state space, c) allows systematic study of how different feature types influence user behavior, and d) remains computationally feasible without explicitly enumerating all state-feature combinations.

\subsection{Contextual Features}

Context is a broad concept that encompasses many factors influencing a user’s decisions during an interactive search session. To capture these influences, we select a diverse set of features representing complementary aspects of context. These features can be broadly grouped into two categories: \textit{situational features}, which describe the user’s state and the progress of the search session, and \textit{content-based features}, which characterize the relevance or quality of the document or snippet currently under consideration.

\subsubsection{Situational features}
Situational features describe the current context of the user within a session and are here further divided into positional, temporal, and knowledge-based features.

Positional and temporal features reflect how far a user has progressed in a search. Indicators such as rank or elapsed time provide information about the depth of interaction and the effort invested so far. 
User behavior is known to be strongly influenced by search progress: rank bias has been described repeatedly, and rank is widely used in information retrieval evaluation as a proxy for user attention~\cite{Joachims2005,Craswell2008,Liu2017}. Similarly, elapsed time is associated with factors such as fatigue, diminishing returns, or the likelihood of abandoning a search~\cite{Jiang2014,Zhang2011}. For these reasons, positional and temporal features are included as fundamental situational signals.

However, progress alone does not capture what a user has learned during a session. Whether a user continues or terminates a search depends not only on how long or how deep they have searched, but on whether their information need has been satisfied and which actions they have already taken~\cite{Thomas2014,White2009,Borisov2016}. 
To model this aspect, we introduce a low-level knowledge-state representation that reflects prior user interactions. In particular, we consider how many documents have already been clicked or, where applicable, explicitly marked as relevant. These signals serve as a proxy for accumulated knowledge and perceived success, allowing the model to account for the influence of prior interactions on subsequent decisions.

\subsubsection{Content\hyp{}based features}
Content\hyp{}based features describe properties of the document or snippet currently under consideration and are independent of the user’s situational state. Examples include a document's retrieval score or the similarity between document text and query. Such features provide an estimate of result quality from the system’s perspective and can influence how users respond to presented items: low-quality results may encourage query reformulation, while high-quality results are more likely to be clicked or to motivate deeper exploration of the current result list~\cite{Abualsaud2019,Azzopardi2020}. By incorporating content-based features, the model accounts for how perceived relevance shapes user decisions at each interaction step.

\subsubsection{Combined feature representations}
In realistic search scenarios, user decisions are rarely driven by a single factor; rather, they emerge from the interaction between search progress, accumulated knowledge, and the perceived quality of available results. Evaluating feature groups individually and in combination allows us to analyze which sources of contextual information contribute to different aspects of behavioral realism and at what levels of behavioral specificity these effects become apparent.

\section{Methodology}

We aim to examine how various types of contextual information influence behavioral realism in Markov-style user models, and how these effects manifest in predictive fit and simulated user behavior in different search settings. To this end, we use three heterogeneous session log datasets and train multiple models on the observed interactions: a baseline Markov model with constant transition probabilities, and feature-conditioned models incorporating different combinations of situational and content-based features.

To evaluate model performance while avoiding overfitting, we employ 10-fold cross-validation with repeated simulation. 

Evaluation is conducted at multiple levels of specificity, ranging from overall aggregated interaction counts to path-level similarity and the preservation of original behaviors such as clicks. To assess the benefit of feature conditioning, we compare the performance of extended models to the baseline. Furthermore, by systematically comparing models with different feature combinations, we identify which types of contextual features contribute most to behavioral realism and map their effects across levels of user interaction.

\subsection{Datasets}

For evaluation, we needed datasets with sufficiently long sessions, the required contextual features, and comparable interaction logging. These characteristics are typically found in controlled laboratory studies, so we selected three lab-study datasets. These cover different search settings, enabling a systematic analysis of contextual effects across heterogeneous scenarios. While two datasets involve conventional search setups\textemdash in news search and argument retrieval\textemdash the third uses predefined queries, resulting in a more constrained search process. Moreover, its results are presented in randomized rather than relevance-ranked order, allowing us to study user behavior in the absence of rank-based relevance cues and to disentangle positional effects from document relevance.

\subsubsection{TREC Session 2014}

The first dataset is based on Session Track data from TREC 2014~\cite{DBLP:conf/trec/CarteretteKHC14}, a widely used and well-established benchmark for modeling user behavior in search, and represents the most common search scenario in this work. The logs were collected via Mechanical Turk on ClueWeb12 and contain interaction data, including initial queries, reformulations, and document clicks. Only implicit relevance feedback is available; there is no information about user satisfaction or explicit relevance judgments.

Of the 1,257 sessions in the dataset, the majority contained too few interactions for an informative analysis. We therefore selected a subset of 200 sessions from the dataset, requiring each session to contain at least two reformulations and three click interactions.

Although interaction timings for queries and clicks are recorded, there is no explicit indication of session termination. Furthermore, while document rankings are available, the original retrieval scores could not be reconstructed. However, similarities between queries and document titles or snippets could still be computed.

The sessions are typically short and limited to basic interaction types, which constrains the expressiveness of user behavior and reduces the observable differences between models. Nevertheless, the dataset remains an important standard benchmark, allowing for comparison with prior work.

\subsubsection{LISP}

The second dataset, LISP~\cite{friese2026lisprichinteraction}, 
consists of session logs from the argument retrieval domain, representing a distinct search task while providing a comparable setup. It contains 122 user sessions from 61 distinct users and was collected in a controlled lab environment. Participants were asked to prepare an opinion essay by searching for pro- and con-arguments for a given topic within a fixed time frame of 10 minutes. During a session, users could reformulate queries, explore result pages, read full documents, and mark arguments as relevant, providing explicit relevance feedback.

In addition to interactions and document information, the logs include timestamps for query formulation, document expansion and reduction, and marking actions; intermediate snippet observations were not explicitly logged and therefore had to be estimated.  
Moreover, the original retrieval pipeline was published with the dataset, enabling us to compute additional text-based features beyond the provided retrieval scores.

Overall, LISP sessions are relatively long and contain a rich set of interactions, offering a detailed view of user behavior.

\subsubsection{TIPPS} 
Finally, we used interaction logs collected in the study \enquote{The Influence of Presentation and Performance on User Satisfaction} (here referred to as TIPPS) by Pathak et al.~\cite{Pathak2024} that were kindly provided on request. The study investigated the effects of result card layouts on user search performance and satisfaction. Participants searched for news on given topics using a set of provided queries. While they could decide which queries to inspect and in what order, they could not formulate their own.

A particularly interesting aspect of this dataset is that search results were presented in randomized order rather than ranked according to system-estimated relevance. Although the dataset addresses a similar news search task as TREC, the randomized result presentation leads to a substantially different interaction setting, which allows us to examine the influence of positional bias and to investigate whether users are more strongly affected by a document’s presented rank or by its actual relevance.

The dataset did not include raw interaction logs but rather preprocessed data from which we reconstructed user sessions. The available data contain clicks, user relevance judgments, both the original and displayed document ranks, and others. The detailed setup description allowed for computation of content-based features, including retrieval scores, though missing document IDs prevented their extraction for some documents in the logs.

We restricted our analysis to sessions originating from one of the evaluated interfaces. Specifically, we selected the interface most comparable to those used in the other datasets with respect to result presentation, ensuring that the same feature sets could be applied reasonably across datasets. This resulted in a final set of 75 sessions.

\subsection{Data Preprocessing and Feature Extraction}
\label{subsec:preprocessing_and_features}

For each dataset, the session logs were preprocessed to extract relevant user interactions and transform the data into a uniform format suitable for modeling. Non-informative entries were removed, and relevant interactions were represented as single records containing all necessary contextual information. Since snippet views were not explicitly recorded, we estimated seen snippets based on clicked documents and the time intervals between consecutive interactions.

After preprocessing, features for the feature-conditioned models were computed where necessary and grouped into related sets:
\begin{itemize}
\item \textit{Positional:} rank, query rank
\item \textit{Knowledge-based:} number of previous clicks, number of previous marks (where available)
\item \textit{Temporal:} time spent in the session
\item \textit{Content-based:} retrieval score (where available), query\hyp{}document/snippet similarity
\end{itemize}

The availability of content\hyp{}based features differed between data\-sets: For TREC, no full texts were available, so query\hyp{}snippet cosine similarity was used instead of query\hyp{}document cosine similarity.

Regarding temporal features, interaction times in both the simulations and the original sessions were set to the average interaction times observed in the dataset. While this normalization leads to the measures being functions of interaction counts and ordering and not independent of other feature sets, it ensures comparability between real and simulated sessions and avoids distortion by user-specific timing effects.

\subsection{Model Training}

For each state, we trained a separate multi-class logistic regression model to predict the next user action. Models were trained for each feature set and for combinations of feature sets.

Training was performed within each fold. 
Observed next actions were encoded as categorical labels, and feature vectors were standardized. 
Hyperparameters, including the regularization strength, were optimized via 5-fold grid search within the training data, using negative log-likelihood as the scoring metric. After optimization, models were retrained on the full training set within the fold.
The baseline Markov model with constant transition probabilities was trained directly from observed transitions, without any contextual features.
This approach ensures that each state’s transition probabilities are learned independently while allowing comparisons between models trained with different feature sets.

\begin{figure}
    \centering
    \includegraphics[width=.97\linewidth]{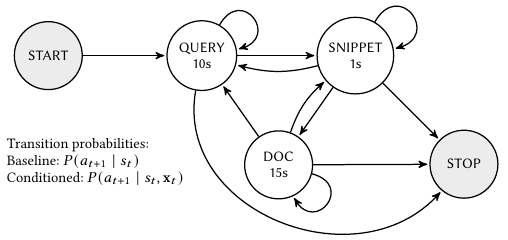}
    \caption{State-based user model used for simulation (TREC configuration). Nodes represent interaction states and directed edges indicate possible transitions. Times per state are estimated averages over the whole dataset. The LISP and TIPPS datasets additionally include a \sffamily{MARK} state.}
    \Description{Image of a Markov model with five states: START, QUERY, SNIPPET, DOC, and END. START transitions to QUERY. QUERY, SNIPPET, and DOC are mutually connected, and each can transition to END. The transition probabilities are described on the left for the baseline and conditioned models. For the baseline model, the transition probability is the conditional probability of $a_{t+1}$ given $s_t$. For the conditioned model, it is the conditional probability of $a_{t+1}$ given $s_t, \mathbf{x}_t$.}
    \label{fig:markov_model}
\end{figure}

Figure~\ref{fig:markov_model} illustrates a model resulting from this process.

\subsection{Simulation Procedure}
\label{subsec:simulation}

To assess behavioral realism, we simulated user sessions based on the original session logs. For each observed session, we ran 100 simulations per model, and outcomes were averaged across runs to reduce Monte Carlo noise (i.e., variability due to stochastic sampling). The procedure was identical for both the baseline and feature-conditioned models. 

Each simulation begins in a dedicated {\sffamily START} state and proceeds until the session reaches a {\sffamily STOP} state, with a fixed interaction time assigned to each state. At each step, the next action is drawn according to the model’s predicted probabilities, conditioned on the current state and, for feature-conditioned models, the current feature vector. Features that depend on session history (e.g., number of documents already clicked, time spent) are updated dynamically during the simulation, and the updated values are used for subsequent transitions.

Queries and their corresponding result lists are taken directly from the original session logs. When a simulated user enters a {\sffamily QUERY} state, the next query and associated ranked documents are retrieved from the corresponding real session. The simulated user then interacts with the result list, clicking or marking documents according to the model’s probabilities. If a query reformulation occurs, the next query is again taken from the original session; if the queries in the session are exhausted, the session terminates at the next reformulation and enters the {\sffamily STOP} state.

As a consequence of this design, simulated sessions can contain at most as many queries as the corresponding real session and are therefore, on average, shorter. We intentionally impose this constraint to ensure that all models are evaluated under identical and realistic conditions. Generating artificial queries would introduce additional modeling assumptions and potential biases that are difficult to quantify, and would remove the direct comparability between simulated and real sessions that underlies our evaluation. By grounding simulations in the original queries and result lists, we preserve a consistent reference point for assessing behavioral fidelity while allowing the models to express different decision-making patterns at each interaction step.

\section{Evaluation}

To assess the quality of the proposed models, we adopt a multi-level evaluation framework. This approach allows us to evaluate not only predictive performance but also behavioral realism at progressively finer levels of detail, from coarse aggregate measures to detailed interaction preservation and session trajectory similarity. 

\subsection{Evaluation Levels}
We evaluate simulated sessions at six levels:

\subsubsection*{1. Predictive Fit}
At the coarsest level, we measure negative log-likelihood (NLL) on held-out sessions. NLL quantifies how well a model predicts the observed actions: lower NLL indicates that the model assigns higher probabilities to the actions that users actually took. This provides a basic check of model fit and ensures that the models are capturing fundamental patterns of decision-making.

\subsubsection*{2. Aggregate Behavior}
Next, we examine session-level statistics to determine whether the models reproduce broad behavioral patterns. We consider:
\begin{itemize}
    \item Number of queries issued
    \item Number of snippets examined
    \item Number of documents clicked
    \item Number of documents marked (when available)
\end{itemize}
For each session, results are averaged over the simulation runs and compared to the corresponding real session. These aggregate measures provide insight into whether the models capture overall interaction volume and search effort.

\subsubsection*{3. Document-level overlap}
To evaluate whether models select the same items as real users, we compute document-level precision, recall, and accuracy for clicks and marks. Accuracy is restricted to the top 30 ranks to avoid inflation from tail-end documents. This analysis captures the models’ ability to reproduce specific document-level decisions, moving beyond coarse aggregate statistics.

\subsubsection*{4. Temporal dynamics}
User behavior is shaped not only by actions, but also by their timing. We measure the following temporal aspects:
\begin{itemize}
    \item time to first click
    \item time to first mark (when available)
    \item time to relevant (TTR)
    \item total session time
\end{itemize}
Time to first click or mark provide straightforward indicators of how quickly users begin interacting with results. TTR captures a more comprehensive notion of efficiency by measuring the average time required to reach relevant documents throughout a session, including interactions that occur later~\cite{Tran2017}. 
As explicit user relevance judgments are available for two datasets, while objective relevance labels are mostly sparse, we use marks to model relevance for LISP and TIPPS, and clicks as implicit relevance signals for TREC.

\subsubsection*{5. Outcome quality}

We compute session\hyp{}level normalized discounted cumulative gain (sDCG)~\cite{sdcg} to evaluate the effectiveness of simulated interactions from a retrieval perspective.

While temporal metrics rely on interaction-based relevance proxies, sDCG is computed using the dataset relevance judgments so that real and simulated sessions can be evaluated on the same scale.

\subsubsection*{6. Trajectory similarity}

Finally, we assess session-level realism by measuring Fréchet distance. Fréchet distance correlates well with established evaluation metrics and is particularly effective for longer and more complex interaction sequences, even when interaction data are sparse~\cite{frechet}. By accounting for both the ordering and the spatial alignment of interactions, it captures similarities in complete behavioral paths rather than isolated actions, complementing more specific measures in this evaluation.

\subsection{Evaluation Strategy}

To assess the quality and behavioral fidelity of the simulated sessions, we adopt a multi-level evaluation strategy that complements the metric definitions. First, each session is simulated 100 times per model to reduce stochastic noise. Metrics are calculated for each simulated session and then averaged across the runs, providing stable per-session estimates.

For each metric, we consider both the raw simulation values and the differences relative to the corresponding real session. This approach allows us to evaluate not only overall behavior but also how closely the model reproduces individual user interactions. After computing per-session measures, values are further averaged across all sessions in the dataset to obtain a single summary value per model for cross-model comparisons.

As this study is exploratory in nature, our goal is not to identify a single statistically superior model, but rather to analyze how different sources of contextual information contribute to simulation realism. We therefore restrict statistical testing to comparisons between each feature-conditioned model and the baseline model. We report paired t-tests with Benjamini--Hochberg FDR correction. In the result tables, statistically significant differences are marked with asterisks (*), while arrows indicate whether the feature-conditioned model produces behavior closer to ($\uparrow$) or further from ($\downarrow$) the real sessions relative to the baseline.

The multi-level evaluation provides several advantages. First, progressive specificity\textemdash from predictive fit to detailed behavioral and trajectory measures\textemdash enables identification of where improvements occur. Second, complementary perspectives allow us to detect cases where models reproduce aggregate statistics well but fail to capture fine-grained document choices or temporal patterns. Third, interpretability is enhanced by separating metrics by level, making it possible to determine which features drive improvements in different aspects of behavior. Finally, robustness is achieved by evaluating across multiple levels and datasets, ensuring that observed gains are not confined to a single metric or dataset.

\section{Results}

In this section, we present the results of our experimental evaluation. Due to space constraints, figures and tables report only the baseline model, the individual feature-set models described in Section~\ref{subsec:preprocessing_and_features}, and the full model combining all feature types, and we focus on mean differences instead of raw values where appropriate. Additional feature combinations were evaluated but are omitted from most visualizations; their outcomes are discussed in the text where they yield notable differences or provide further insight.

\subsubsection*{Predictive Fit}

\begin{figure}[b]
    \centering
    \begin{subfigure}[a]{\linewidth}
    \includegraphics[width=.97\linewidth]{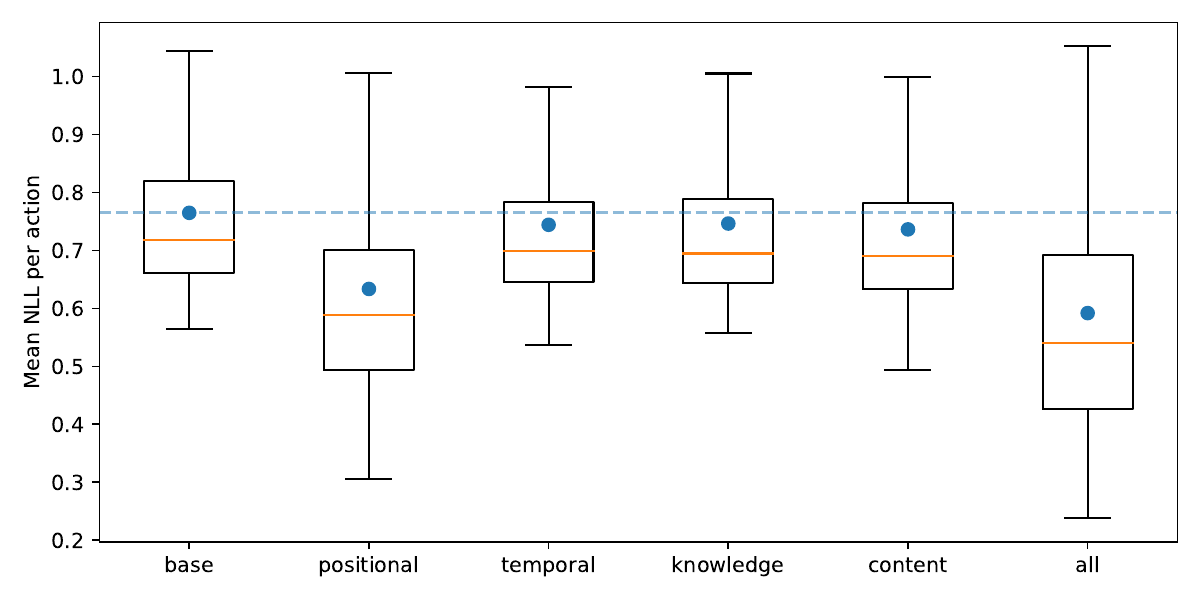}
    \caption{TREC data}
    \label{fig:nll_trec}
    \end{subfigure}\hfill    
    \begin{subfigure}[b]{\linewidth}
    \includegraphics[width=.97\linewidth]{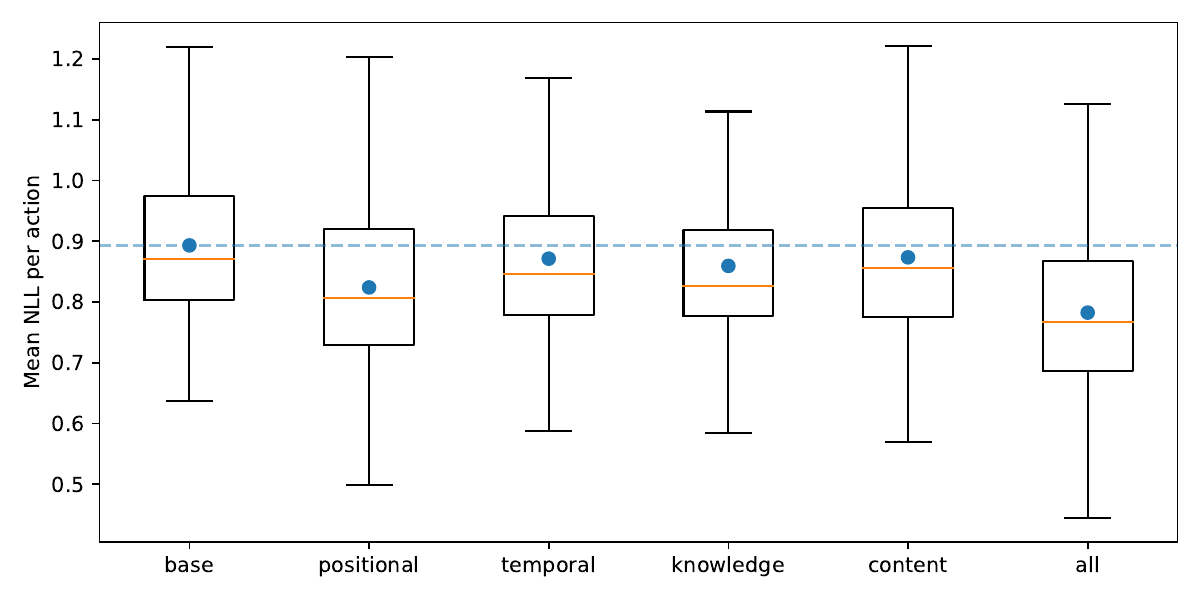}
    \caption{LISP data}
    \label{fig:nll_lisp}
    \end{subfigure}\hfill
    \begin{subfigure}[b]{\linewidth}
    \includegraphics[width=.97\linewidth]{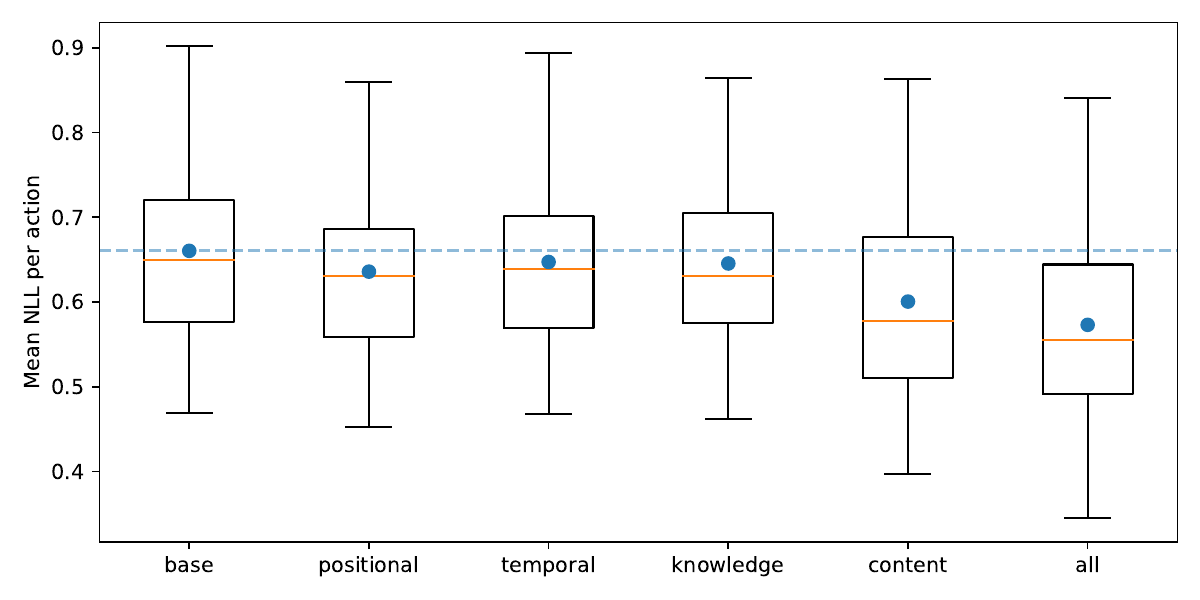}
    \caption{TIPPS data}
    \label{fig:nll_ds3}
    \end{subfigure}
    \caption{Distribution of per-session mean NLL per step for each model. Boxes indicate interquartile ranges, horizontal lines show medians, and dots denote means. Lower values indicate better predictive fit. The dashed blue line represents the mean of the baseline model for comparison.}
    \Description{Three vertically stacked box-plots show the distribution of per-session mean negative log-likelihood (NLL) per step for six models: base, positional, temporal, knowledge, content, and all, one plot per datasets. In each panel, interquartile ranges, medians, ranges, and means are depicted. A dashed blue horizontal line marks the mean of the baseline model. Lower NLL values indicate better predictive fit. For all datasets, all means are below the baseline mean, and all has the lowest mean. For TREC and LISP, positional is the second best; for TIPPS, it is content. Most improvements over the baseline are small.}
    \label{fig:nll}
\end{figure}

Across all datasets, feature-conditioned models consistently achieve better predictive fit than the baseline model (see Figure~\ref{fig:nll}). Although the magnitude of improvement varies, every contextual variant reduces average per-action NLL relative to the baseline. The strongest performance is obtained by the full model that combines all feature types, reaching an average per-action NLL of 0.59 on TREC, 0.78 on LISP, and 0.57 on TIPPS, compared to baseline values of 0.76, 0.89, and 0.66, respectively.

\begin{table*}
\centering
\caption{Mean per-session deviation between simulated and original sessions for aggregate interaction counts. Values report the average difference from the corresponding real sessions; lower values indicate closer behavioral fidelity. \textit{Original} shows the mean counts in the real data for reference.}
\label{tab:counts}
\begin{tabular}{lS[table-format=1.2]S[table-format=1.2]S[table-format=1.2]|S[table-format=1.2]S[table-format=1.2]S[table-format=1.2]|S[table-format=1.2]S[table-format=1.2]S[table-format=1.2]|S[table-format=1.2]S[table-format=1.2]}
\toprule
Model & \multicolumn{3}{c|}{$\Delta$Queries} & \multicolumn{3}{c|}{$\Delta$Snippets} & \multicolumn{3}{c|}{$\Delta$Clicks} & \multicolumn{2}{c}{$\Delta$Marks} \\
& {TREC} & {LISP} & {TIPPS} & {TREC} & {LISP} & {TIPPS} & {TREC} & {LISP} & {TIPPS} & {LISP} & {TIPPS} \\

\cline{1-12}
\textit{Original} & \itshape3.35 & \itshape3.39 & \itshape3.43 & \itshape23.51 & \itshape64.61 &\itshape53.55& \itshape4.21 & \itshape21.90 & \itshape14.59 & \itshape12.58 & \itshape8.56 \\
\cline{1-12}
Baseline & 0.93 & 1.33 & 0.79 & \bfseries 11.71 & 25.96 & \bfseries28.00 & 2.11 & 8.80 & 7.59 & 5.06 & 4.43 \\
\hdashline
Positional & 0.67*$^{\up}$ & 1.23*$^{\up}$ & \bfseries 0.41*$^{\up}$ & 12.45*$^{\down}$ & 25.80 & 28.73 & \bfseries 1.32*$^{\up}$ & 6.64*$^{\up}$ & \bfseries 6.98 & 3.62*$^{\up}$ & \bfseries 3.83*$^{\up}$ \\
Temporal & \bfseries 0.55*$^{\up}$ & \bfseries 1.15*$^{\up}$ & 0.51*$^{\up}$ & 13.85*$^{\down}$ & \bfseries 23.03*$^{\up}$ & 31.16*$^{\down}$ & 2.14 & 6.98*$^{\up}$ & 7.69 & 3.76$^{\up}$* & 4.08*$^{\up}$ \\
Knowledge & 0.61*$^{\up}$ & 1.18*$^{\up}$ & 0.53*$^{\up}$ & 14.07*$^{\down}$ & 23.13*$^{\up}$ & 31.26*$^{\down}$ & 2.33*$^{\down}$ & 6.99*$^{\up}$ & 7.94 & 4.42*$^{\up}$ & 4.22 \\
Content & 0.84*$^{\up}$ & 1.31 & 0.73*$^{\up}$ & 15.06*$^{\down}$ & 26.23 & 32.72*$^{\down}$ & 2.63*$^{\down}$ & 9.12 & 9.13*$^{\down}$ & 5.27 & 5.43*$^{\down}$ \\
All & \bfseries 0.55*$^{\up}$ & 1.28 & 0.44*$^{\up}$ & 12.21*$^{\down}$ & 24.12*$^{\up}$ & 29.59 & 1.75*$^{\up}$ & \bfseries 5.42*$^{\up}$ & 7.54 & \bfseries 3.42*$^{\up}$ & 4.07 \\
\bottomrule
\end{tabular}
\end{table*}

Among the individual feature sets, positional features yield the most substantial improvements for TREC and LISP, with NLLs of 0.63 on TREC and 0.82 on LISP, already representing a clear increase in predictive fit over the context-free model. Other single-feature models provide only slight benefits. On TIPPS, the content-based model shows the strongest improvement over the baseline with an average per-action NLL of 0.60.

Combining multiple feature sets generally leads to further improvements, but the effect is incremental: pairwise combinations result in only minor gains over the best individual models, while more pronounced improvements emerge only when three feature types are integrated. Even then, none of the partial combinations match the performance of the full feature set.

\subsubsection*{Aggregate behavior}
At the aggregate level, we compare simulated and real sessions in terms of interaction volume. Rather than absolute counts, Table~\ref{tab:counts} reports the mean per-session deviation between simulated and real sessions for each interaction type, reflecting how closely a model reproduces overall user activity.

Across datasets, all models underestimate interaction volume relative to the original logs, which is expected given that simulations are constrained by observed query sequences. Aggregate deviations should therefore be interpreted in light of the simulation setup, as early termination directly affects interaction counts and session duration. 
As a result, differences at this level partly reflect structural truncation rather than purely behavioral modeling quality. Nevertheless, systematic differences between models are apparent.

Overall, feature-conditioned models reduce aggregate deviations compared to the baseline, although no single model consistently performs best across all interaction types and datasets. For all datasets, contextual conditioning improves the accuracy of issued query counts, with the temporal model achieving the lowest query-level deviation for TREC and LISP. For LISP, temporal features yield comparatively strong performance across several aggregate measures, whereas for TREC, their benefits are largely limited to query counts, and for TIPPS, they do not show substantial benefits at all. 

For snippet interactions, results differ across datasets. On TREC and TIPPS, none of the single-feature models outperforms the baseline, although the combination of positional and knowledge features yields the smallest deviation for TREC, with an average difference of 10.92 snippets per session. On LISP, however, most feature-conditioned models reduce deviations relative to the baseline.

Click and mark interactions show more pronounced improvements, though not for all feature sets. Positional features and the full feature combination consistently yield the smallest deviations for clicks and marks across all datasets. In contrast, content-based features do not improve aggregate interaction volumes on any dataset, while temporal and knowledge features lead to modest improvements on LISP but not on TREC or TIPPS.

Taken together, positional features exhibit a strong and consistent performance profile: while not always yielding the smallest deviations, they reliably outperform the baseline and perform well on the most behaviorally salient measures, such as clicks and marks. The full feature combination generally achieves comparable results with at most small gains over positional features.

Notably, certain intermediate feature combinations perform particularly well on specific measures. The combination of positional and knowledge features yields the smallest deviations for snippet interactions on TREC and for clicks and marks on LISP (mean differences of 1.37 and 2.70, respectively), as well as for query counts on TREC (0.49). On TIPPS, the knowledge-temporal-combination model achieves the best result for number of marked documents with a mean per-session deviation of 3.66. However, these combinations perform mostly below average on other measures, highlighting trade-offs between feature types.

\subsubsection*{Document-level overlap}

At the document level, feature\hyp{}conditioned models consistently outperform the baseline on TREC and LISP (see Table~\ref{tab:doc_overlap}). Precision gains are modest, suggesting limited ability to select only the documents users actually chose. In contrast, recall (i.e., click or mark preservation) and accuracy improve more substantially, indicating that feature conditioning helps recover more of the documents users interacted with, particularly among top-ranked results where attention is concentrated.

\begin{table*}
\centering
\caption{Document-level overlap between simulated and real sessions, measured by precision, recall, and accuracy@30. Higher values indicate closer agreement with the original user interactions.}
\label{tab:doc_overlap}
\begin{tabular}{lS[table-format=1.2]|S[table-format=1.2]S[table-format=1.2]|S[table-format=1.2]S[table-format=1.2]|S[table-format=1.2]|S[table-format=1.2]S[table-format=1.2]|S[table-format=1.2]S[table-format=1.2]|S[table-format=1.2]|S[table-format=1.2]S[table-format=1.2]|S[table-format=1.2]S[table-format=1.2]}
\toprule
Model & \multicolumn{5}{c|}{Precision} & \multicolumn{5}{c|}{Recall} & \multicolumn{5}{c}{Accuracy}  \\
& {TREC} & \multicolumn{2}{c|}{LISP} & \multicolumn{2}{c|}{TIPPS} & {TREC} & \multicolumn{2}{c|}{LISP} & \multicolumn{2}{c|}{TIPPS} & {TREC} & \multicolumn{2}{c|}{LISP} & \multicolumn{2}{c}{TIPPS} \\
& {click}  & {click} & {\hspace{2pt}mark\hspace{2pt}} & {click} & {\hspace{2pt}mark\hspace{2pt}} & {click} & {click} & {\hspace{2pt}mark\hspace{2pt}} & {click} & {\hspace{2pt}mark\hspace{2pt}} & {click} & {click} & {\hspace{2pt}mark\hspace{2pt}} & {click} & {mark} \\
\cline{1-16}

Baseline    & 0.09 & 0.40 & 0.21 & 0.30 & 0.15 & 0.31 & 0.33 & 0.28 & 0.18 & 0.21 & 0.79 & 0.65 & 0.77 & 0.66 & 0.76 \\
\hdashline
Positional  & 0.13*$^{\up}$ & \bfseries0.42*$^{\up}$ & \bfseries0.24*$^{\up}$ & 0.24*$^{\down}$ & 0.14*$^{\down}$ & 0.38*$^{\up}$ & 0.44*$^{\up}$ & \bfseries0.35*$^{\up}$ & 0.18 & 0.24*$^{\up}$ & 0.84*$^{\up}$ & 0.77*$^{\up}$ & 0.82*$^{\up}$ & 0.78*$^{\up}$ & 0.83*$^{\up}$ \\
Temporal    & 0.10 & 0.41*$^{\up}$ & 0.23*$^{\up}$ & 0.25*$^{\down}$ & 0.15 & 0.33*$^{\up}$ & 0.38*$^{\up}$ & 0.33*$^{\up}$ & 0.17 & 0.23*$^{\up}$ & 0.85*$^{\up}$ & 0.77*$^{\up}$ & 0.82*$^{\up}$ & 0.77*$^{\up}$ & 0.83*$^{\up}$ \\
Knowledge   & 0.09 & 0.40 & 0.22*$^{\up}$ & 0.25*$^{\down}$ & 0.14 & 0.32*$^{\up}$ & 0.38*$^{\up}$ & 0.31*$^{\up}$ & 0.16*$^{\down}$ & 0.23*$^{\up}$ & 0.85*$^{\up}$ & 0.76*$^{\up}$ & 0.82*$^{\up}$ & 0.77*$^{\up}$ & 0.83*$^{\up}$ \\
Content     & 0.11*$^{\up}$ & 0.41*$^{\up}$ & 0.22*$^{\up}$ & \bfseries0.33*$^{\up}$ & \bfseries0.20*$^{\up}$ & 0.33*$^{\up}$ & 0.36*$^{\up}$ & 0.30*$^{\up}$ & 0.19 & 0.24*$^{\up}$ & \bfseries0.86*$^{\up}$ & \bfseries0.78*$^{\up}$ & \bfseries0.83*$^{\up}$ & 0.79*$^{\up}$ & \bfseries0.84*$^{\up}$ \\
All         & \bfseries0.14*$^{\up}$ & 0.41*$^{\up}$ & 0.23*$^{\up}$ & 0.30 & \bfseries0.20*$^{\up}$ & \bfseries0.39*$^{\up}$ & \bfseries0.45*$^{\up}$ & \bfseries0.35*$^{\up}$ & \bfseries0.23*$^{\up}$ & \bfseries0.29*$^{\up}$ & 0.85*$^{\up}$ & 0.76*$^{\up}$ & 0.81*$^{\up}$ & \bfseries0.80*$^{\up}$ & \bfseries0.84*$^{\up}$ \\
\bottomrule
\end{tabular}
\end{table*}

For TIPPS, where the result lists are not ordered by relevance, several models show little to no improvement over the baseline. By removing the implicit relevance signal that is typically conveyed through document rank, the randomized presentation makes it difficult for models without access to document content to reliably predict whether a document will be clicked.
Accordingly, the content-based model is the best-performing standalone model on this dataset, achieving the strongest overall results for click precision, mark precision, and mark accuracy. Combining content-based and positional features yields further improvements, resulting in performance that nearly matches that of the full feature set model, which achieves the best overall results on this dataset. 

Across the other datasets and interaction types, positional features yield the strongest single-feature improvements. On TREC, recall increases from 0.31 to 0.38 and accuracy from 0.79 to 0.84, while on LISP clicks, recall improves from 0.33 to 0.44 and accuracy from 0.65 to 0.77. For marks in LISP, similar gains are observed. Adding further feature types produces no notable benefits, and the combined model performs comparably to the positional model rather than surpassing it. Other feature sets\textemdash temporal, knowledge-based, and content-based\textemdash achieve small but consistent improvements over the baseline regarding precision and recall, yet remain below the positional model for these datasets. In terms of accuracy, all feature-conditioned models show similar performances.

\subsubsection*{Temporal dynamics}
Table~\ref{tab:temporal} reports mean deviations between simulated and real sessions for the time-based measures: time to relevant (TTR), time to first click/mark, and total session duration. As a result of the temporal normalization described in Section~\ref{subsec:preprocessing_and_features}, times reported in the row \textit{Original} only approximate the observed values. Moreover, like the aggregate counts in Table~\ref{tab:counts}, session times are biased by the experimental setup due to the simulation constraints.

At this level, the baseline model exhibits a strong performance: for several measures on the LISP and TIPPS datasets, it ranks among the best configurations, contrasting with earlier evaluation levels.

Also in contrast to the previous levels, positional features perform relatively poorly here for TREC and LISP, but notably well for TIPPS. While positional conditioning helps predict which documents users interact with in ordered search settings, it does not appear to capture the timing of these interactions with comparable fidelity. However, it provides the most reliable signal when relevant document placement within the result list is uncertain.

For TREC, temporal features expectedly yield strong improvements regarding time-based measures, particularly for TTR and time to first click, which is consistent with their design. More notably, knowledge- and content-based features also perform very well on these metrics, often matching or exceeding the performance of the temporal model. The combination of knowledge and content features achieves the lowest deviations for both TTR and time to first click, with mean differences of 0.07 and 0.12, respectively. 
On LISP, temporal features do not show improvements over the baseline. Instead, knowledge-based and content-based models achieve the lowest deviations among single-feature configurations for several measures, including TTR and time to first click. As observed for TREC, models that showed comparatively weaker performance on aggregate and document-level metrics exhibit stronger performance on temporal aspects on LISP. 

On TIPPS, temporal features show uneven effects, with clear improvements over the baseline for some measures but very pronounced disadvantages for others.

The full feature combination is not among the best models at this level, only leading for overall session time on LISP and frequently underperforming the baseline. 

\begin{table*}
\centering
\caption{Temporal characteristics of simulated sessions compared to original sessions. Values report mean differences for time to relevant (TTR), time to first click, time to first mark, and total session duration. Smaller values indicate closer agreement with observed user behavior.}
\label{tab:temporal}
\begin{tabular}{lS[table-format=1.2]S[table-format=1.2]S[table-format=1.2]|S[table-format=1.2]S[table-format=1.2]S[table-format=1.2]|S[table-format=1.2]S[table-format=1.2]|S[table-format=1.2]S[table-format=1.2]S[table-format=1.2]}
\toprule
Model & \multicolumn{3}{c|}{$\Delta$TTR} & \multicolumn{3}{c|}{$\Delta$1st Click} & \multicolumn{2}{c|}{$\Delta$1st Mark} & \multicolumn{3}{c}{$\Delta$Session} \\
& {TREC} & {LISP} & {TIPPS} & {TREC} & {LISP} & {TIPPS} & {LISP} & {TIPPS} & {TREC} & {LISP} & {TIPPS} \\
\cline{1-12}
\textit{Original} & \textit{15.27} & \textit{43.25} & \textit{8.04} & \textit{20.88} & \textit{19.42} &\textit{17.31}& \textit{48.78} & \textit{48.01} & \textit{126.19} & \textit{445.57} & \textit{321.17} \\
\cline{1-12}
Baseline    & 2.19 & 6.31 & 3.34 & 2.17 & \textbf{0.68} & 2.00 & 2.55 & \textbf{1.13} & 53.05 & 176.80 & 154.79 \\
\hdashline
Positional  & 4.19*$^{\down}$ & 8.55*$^{\down}$ & 1.91*$^{\up}$ & 3.26*$^{\down}$ & 2.17*$^{\down}$ & \bfseries0.54*$^{\up}$ & 5.83*$^{\down}$ & 6.55*$^{\down}$ & \bfseries39.52*$^{\up}$ & 141.78*$^{\up}$ & \bfseries142.04*$^{\up}$ \\
Temporal    & 0.56*$^{\up}$ & 8.91*$^{\down}$ & 2.98*$^{\up}$ & 0.29*$^{\up}$ & 1.49*$^{\down}$ & 0.96*$^{\up}$ & 7.34*$^{\down}$ & 7.56*$^{\down}$ & 55.71 & 143.34*$^{\up}$ & 156.43 \\
Knowledge   & \bfseries0.08*$^{\up}$ & \bfseries5.16*$^{\up}$ & 2.10*$^{\up}$ & \bfseries0.19*$^{\up}$ & 1.45*$^{\down}$ & 1.11*$^{\up}$ & 3.19 & 5.88*$^{\down}$ & 55.71*$^{\down}$ & 144.66*$^{\up}$ & 160.57 \\
Content     & 0.98*$^{\up}$ & 5.91 & 2.63 & 0.55*$^{\up}$ & 0.70 & 1.72 & \textbf{2.30} & 1.74 & 63.40*$^{\down}$ & 181.73 & 182.99*$^{\down}$ \\
All         & 2.88*$^{\down}$ & 5.81 & \bfseries 0.96*$^{\up}$ & 2.61*$^{\down}$ & 1.97*$^{\down}$ & 0.85*$^{\up}$ & 4.34*$^{\down}$ & 4.45*$^{\down}$ & 44.48*$^{\up}$ & \bfseries122.11*$^{\up}$ & 151.84 \\
\bottomrule
\end{tabular}
\end{table*}

\subsubsection*{Outcome quality}
Across datasets, the full model achieves the lowest sDCG deviation from the original sessions, indicating the most faithful reproduction of retrieval outcomes (see Table~\ref{tab:sdcg_and_frechet}). However, the combined model offers only minimal improvement over the strongest single feature set on TREC and LISP, where positional features already substantially outperform the baseline. Only on TIPPS we see a substantial improvement by combining feature sets.

On LISP, all feature-conditioned models clearly outperform the baseline. On TREC, improvements are less uniform: while the positional and combined models provide clear benefits, other feature sets yield only minor gains relative to the baseline. Nevertheless, the relative ordering of models is consistent across TREC and LISP, with positional and combined configurations performing best and the remaining feature sets showing similar, intermediate performance.

Again, the different setup of the TIPPS study is reflected in a different ordering of models. Aligning with earlier levels, the content and positional models present the best single feature set options. However, for substantial improvements at this level, feature sets have to be combined. Combining the two top performing models reduces the mean deviation in sDCG to 0.11, almost achieving the top result of the maximum combination model of 0.10.

\begin{table}[t]
\centering
\caption{Session-level outcome quality measured by mean deviation in sDCG (left) and overall trajectory similarity measured by Fréchet distance (right) between simulated and real sessions. Lower values indicate better alignment with real sessions for both metrics.}
\label{tab:sdcg_and_frechet}
\begin{tabular}{l|S[table-format=1.2]S[table-format=1.2]S[table-format=1.2]|S[table-format=1.2]S[table-format=1.2]S[table-format=1.2]}
\toprule
& \multicolumn{3}{c|}{$\Delta$sDCG} & \multicolumn{3}{c}{Fréchet dist.} \\
Model & {TREC} & {LISP} & {TIPPS} & {TREC} & {LISP} & {TIPPS}  \\
\cline{1-7}
\textit{Original}   & \textit{0.38} & \textit{0.34} & \textit{0.57} & {--} & {--} & {--} \\
\cline{1-7}
Baseline   & 0.23 & 0.16 & 0.20 & 1.31 & 9.68 & 2.06 \\
\hdashline
Positional & 0.13*$^{\up}$ & 0.09*$^{\up}$ & 0.16*$^{\up}$ & 0.98*$^{\up}$ & 7.50*$^{\up}$ & 1.67*$^{\up}$ \\
Temporal   & 0.20*$^{\up}$ & 0.12*$^{\up}$ & 0.19 & \bfseries0.76*$^{\up}$ & 7.39*$^{\up}$ & 1.89*$^{\up}$ \\
Knowledge  & 0.22*$^{\up}$ & 0.13*$^{\up}$ & 0.20 & 0.77*$^{\up}$ & 7.68*$^{\up}$ & 1.86*$^{\up}$ \\
Content    & 0.21*$^{\up}$ & 0.13*$^{\up}$ & 0.16*$^{\up}$ & 1.04*$^{\up}$ & 8.55*$^{\up}$ & 1.82*$^{\up}$ \\
All        & \bfseries0.12*$^{\up}$ & \bfseries0.08*$^{\up}$ & \bfseries0.10*$^{\up}$ & \bfseries0.76*$^{\up}$ & \bfseries7.03*$^{\up}$ &
\bfseries1.58*$^{\up}$  \\
\bottomrule
\end{tabular}
\end{table}

\subsubsection*{Trajectory similarity}

Across all datasets, all feature-conditioned models substantially reduce Fréchet distance relative to the baseline, indicating more realistic interaction trajectories. The vast differences between the datasets likely reflect the greater length and complexity of LISP sessions. 
Among the models shown in Table~\ref{tab:sdcg_and_frechet}, the full model achieves the lowest distances overall, yielding the closest match to real session paths. However, the strongest configurations are not always the most complex ones: on TREC, the knowledge–positional combination performs best (FD = 0.71).

\section{Discussion}

Across evaluation levels, feature-conditioned models generally outperform the static baseline, supporting the general expectation behind RQ1. Benefits are particularly evident in predictive fit, document\hyp{}level overlap, session-level outcome measures, and trajectory-level similarity. While improvements are mostly moderate, reflecting both the inherent variability of user behavior and the limitations of Markov-style abstractions, the pattern across datasets and metrics demonstrates the practical benefits of incorporating contextual features into state-based models.

At the same time, the results demonstrate that feature conditioning is not universally beneficial and can, in some cases, even reduce modeling and simulation quality. Contextual features therefore only improve simulation quality when they are genuinely informative of user decisions, making careful feature selection and consideration of feature interactions essential.

Addressing RQ2, different types of contextual features contribute unevenly across behavioral dimensions. Positional features emerge as a strong and robust signal, reliably enhancing document choice and outcome-oriented measures\textemdash both of which are particularly relevant indicators of simulation realism, as they directly affect or pose downstream evaluation outcomes\textemdash in rank-ordered search settings. This finding aligns with established evidence that rank strongly guides user attention and interactions~\cite{10.1145/1229179.1229181,Granka2004}.

At the same time, the TIPPS dataset demonstrates that this effect depends critically on the relationship between rank and relevance. In conventional search settings, rank implicitly conveys information about document quality because results are ordered by estimated relevance. When this assumption is removed through randomized presentation, positional models can still reproduce coarse interaction trajectories, but struggle to identify the specific documents users engage with. This highlights the importance of incorporating document-based features: once rank no longer serves as a proxy for relevance, interactions become more strongly driven by the actual content relevance of documents to the user’s information need.

Beyond the particularly strong impact of content features on TIPPS, content- and knowledge-based features contribute more strongly to temporal dynamics and trajectory-level similarity. This suggests that result quality provides an important cue beyond positions and reinforces that no single feature type suffices to capture all aspects of user behavior. 
However, though clearly beneficial at these levels, content-based models underperform the baseline on most aggregate count metrics. Since content-related features primarily affect local click or mark decisions rather than higher-level behaviors such as session continuation or stopping, no major improvements were expected here. 
Especially in the shuffled setting, where both relevant and non-relevant documents may appear at any rank, predicting how or whether a session will continue based solely on document properties becomes highly difficult without additional search context.
Still, it should be noted that this effect is amplified by the constrained simulation setup: restricting simulations to observed query sequences limits the extent to which local improvements can accumulate into session-level differences, potentially masking their impact through early termination. 

Combined models often perform best by leveraging the strengths of complementary feature types. 
However, indiscriminately including features can harm performance. Predictive fit increases with more features, since NLL, the evaluation criterion at this level, is also the model tuning criterion. Comparable to pointwise learning-to-rank, the models optimize transition probabilities at each step with regard to NLL. But this pattern does not extend to evaluation criteria not used for model tuning, as seen for temporal measures. Based on path-length, these are more comparable to listwise learning-to-rank. Improvements at one level therefore do not necessarily translate to improvements at the other, emphasizing the need for careful feature selection: While contextual features are generally beneficial, their impact depends on alignment with the behavioral dimension under consideration. Selecting complementary features can substantially enhance realism without greatly increasing model complexity, whereas poorly aligned or excessive features may yield diminishing or even negative returns. This highlights a practical guideline for designing feature-conditioned user simulators: more context is not inherently better, and attention to which features support which behavioral outcomes is critical, particularly for simulations intended to inform system-oriented evaluation.

In addition to feature selection, the effectiveness of contextual conditioning also depends on session characteristics. 
While more structured sessions are generally easier to simulate and benefit from the pattern recognition capabilities of feature-conditioned models, less consistent search settings can benefit even more from carefully chosen conditioning features\textemdash while also being more sensitive to unsuitable feature choices.
Besides, longer sessions, such as those in the LISP dataset, show clearer benefits of feature-conditioned models, as longer sessions provide more opportunities for model behavior to diverge, while shorter sessions, like TREC, constrain interaction volumes and limit differentiation between models. This becomes especially apparent at the aggregate count level, where the effect is likely amplified by the simulation constraint described in Section~\ref{subsec:simulation}. This suggests that the advantages of contextual conditioning become more pronounced as sessions grow longer and interaction dynamics unfold over time.

The varying magnitude and consistency of improvements across metrics, as discussed for content-related features above, highlight the importance of a multi-level evaluation perspective. The benefits of contextual conditioning are not uniform across behavioral dimensions: while document choice, session effectiveness, and trajectory similarity clearly benefit from it, the picture is more heterogeneous for predictive fit, aggregate counts, and temporal aspects. Moreover, rankings of models vary across different measures. These patterns reinforce the value of assessing user simulation at multiple levels to obtain a comprehensive picture of model realism.

Taken together, our findings show that contextual features affect search in various ways, and that contextual conditioning improves user simulation realism, but that its benefits depend on both the behavioral dimension under consideration and the choice of features.

\section{Limitations}

While our study demonstrates the benefits of feature-conditioned state-based user models, some limitations must be acknowledged. First, we only evaluated our approach on three datasets, all of moderate size. Although the results are consistent across these datasets, generalizability to other domains, larger-scale logs, or different interaction scenarios remains to be tested.

Second, due to the exploratory nature of this work and the consequently limited scope of statistical testing, our conclusions rely primarily on consistent trends observed across metrics and datasets rather than on statistically confirmed effects, and should be interpreted accordingly.

Third, many of the features considered in this study are highly abstracted. For example, the knowledge signal is a simple count of previously clicked or marked items, and content-related features reflect only limited aspects of result relevance. These simplifications may affect both the relative performance of models and the absolute realism achieved. 

Finally, our simulations rely on real session query sequences, which limits the length and variability of simulated sessions. While this choice ensures realistic grounding, it also 
restricts the expression of long-term behavioral effects, particularly for feature types whose influence is primarily local.
Exploring approaches that can generate plausible novel queries or extend session structures in a controlled manner could provide additional insight into simulation fidelity and allow testing more diverse interaction scenarios.

\section{Conclusion and Outlook}

We introduced a feature-conditioned extension of state-based user models that enables context-aware decision making while preserving their interpretability and structural simplicity. 
Beyond the scope of conventional search methodologies, this framework can be regarded as a comprehensive approach to modeling sequential, context-dependent interactions, thereby indicating its relevance to emerging domains such as conversational search.

Rather than proposing a single ``best'' model, we used this framework to systematically study how different sources of contextual information affect the realism of simulated search behavior: 
Across multiple datasets and evaluation levels, incorporating contextual features frequently improves predictive fit and behavioral fidelity over a context-free baseline. 
Positional features provide the most consistent gains, while other feature types contribute more selectively across behavioral levels. Content-based features, in particular, show no benefits for aggregate session counts\textemdash partly due to the simulation setup\textemdash but substantially improve the accuracy of click and mark decisions and the modeling of temporal dynamics in ways not captured by positional information alone. 
Combining features, when done deliberately, further increases simulation realism: even a small number of carefully chosen features can yield notable improvements in behavioral fidelity. 
However, the effectiveness of contextual features depends on the search setting and modeling objective. Our proposed framework enables the identification of appropriate feature configurations, thereby maximizing the benefits of feature conditioning while avoiding potential drawbacks by supporting informed, task-dependent feature selection.

Methodologically, our study demonstrates the value of a multi-level evaluation framework, which captures both predictive accuracy and behavioral realism across aggregate, document-level, temporal, and trajectory-based dimensions. 

Overall, this work shows that even relatively simple, context-aware extensions of state-based models can enhance the realism of user simulations.
This suggests a practical use case for constructing and applying realistic user 
models grounded in real world session logs.
By learning model parameters from these session logs, the proposed models make behavioral assumptions explicit and provide an empirical foundation for evaluating the effects of various facets (e.g., alternative retrieval models, automated query (re)formulation). Compared to existing simulation approaches that rely on implicit or arbitrary behavioral assumptions, this enables more realistic and reproducible evaluations of interactive search systems.

\begin{acks}
This work is partially funded by Deutsche Forschungsgemeinschaft
(DFG) under grant number 509543643 and within the funding pro-
gramme FH-Personal (PLan CV, reference number 03FHP109) by
the German Federal Ministry of Research, Technology and Space (BMFTR) and Joint Science Conference (GWK). 

We would like to thank Saber Zerhoudi for providing the original code for the Fréchet distance from~\cite{frechet}.
\end{acks}

\begingroup
\renewcommand{\acksname}{GenAI Usage Disclosure}
\begin{acks}
GenAI tools were used solely to polish the writing of this paper. No GenAI tools were used for the preparation, implementation, evaluation, or any other content of this work.
\end{acks}
\endgroup

\bibliographystyle{ACM-Reference-Format}
\balance
\bibliography{bib}


\end{document}